# A model-based approach towards accelerated process development: A case study on chromatography

*Steven Sachio[1,2], Cleo Kontoravdi[1,2], and Maria M. Papathanasiou[1,2]\**

[1]Sargent Centre for Process Systems Engineering, Imperial College London, SW7 2AZ, UK

[2]Department of Chemical Engineering, Imperial College London, SW7 2AZ, UK

Email: maria.papathanasiou11@imperial.ac.uk

## Abstract

Process development is typically associated with lengthy wet-lab experiments for the identification of good candidate setups and operating conditions. In this paper, we present the key features of a model-based approach for the identification and assessment of process design space (DSp), integrating the analysis of process performance and flexibility. The presented approach comprises three main steps: (1) model development & problem formulation, (2) DSp identification, and (3) DSp analysis. We demonstrate how such an approach can be used for the identification of acceptable operating spaces that enable the assessment of different operating points and quantification of process flexibility. The above steps are demonstrated on Protein A chromatographic purification of antibody-based therapeutics used in biopharmaceutical manufacturing.

# 1 Introduction

The rising demand for biopharmaceuticals is dictating the advancement of current processes towards more efficient operation [1-3]. This comes with challenges for manufacturers who target multiple competing Key Performance Indicators (KPIs), such as product quality and process eco-efficiency. In (bio-)pharmaceutical manufacturing, product quality, with respect to efficacy and purity, is a necessary attribute to be met for regulatory authorities to grant market authorization [4]. Similarly, for existing products, manufacturers are required to continuously demonstrate that product quality is satisfied across all released batches. Traditionally, quality specifications are confirmed through a Quality by Testing (QbT) approach, testing the product exhaustively before batch release. Although compliant, QbT can result in high experimentation costs and waiting times, delaying the market release. Furthermore, such an approach does not allow manufacturers to take in-process corrective actions, often leading to lost batches. This can negatively impact the resilience of the end-to-end supply chain and may lead to increased process waste and financial losses [5, 6].

To this end, Quality by Design (QbD) offers a structured perspective, whereby the target product profile, comprising the comprehensive set of attributes and their target levels, is embedded in the design of the process itself. QbD was originally described in the International Conference on Harmonization guidelines (ICH Q8 [7], ICH Q9 [8], ICH Q10 [9], and ICH Q11 [10]) and is increasingly endorsed by regulators as a methodology that offers process flexibility and, importantly, demonstrates control over the process. A key advantage in QbD-supported initiatives is the identification of an operating space, defined by good candidate condition sets, within which the process is guaranteed to meet the target specifications. This is known as Design Space (DSp) [11, 12] and offers greater operational flexibility, as opposed to point-specific operation.

**Quality by Digital Design (QbDD)**

Computer-aided tools have the potential to drive the (bio-)pharmaceutical industry away from purely wet-lab DSp determination, which is cost-intensive and inefficient [11, 13], expediting process development and decreasing the use of solvents and raw materials, thus improving environmental footprint [14, 15]. The use of mathematical models in this industry is a long-standing research area, with most studies focusing on the development and application of mathematical models for process simulation, optimization, and DSp identification [16-25].

The combination of computer-aided tools and QbD principles offers can accelerate process development via high throughput *in silico* experimentation. The combination of computer aided QbD is defined as "Quality by Digital Design" (QbDD) and can be viewed as a branch of QbD, whereby process design and operation decisions are supported by computer models and tools that inform process development and guide wet-lab validation and testing [26]. Specifically, when referring to QbDD as a workflow (Figure 1), we consider that a process starts with the identification of the Target Product Profile (TPP) that refers to the specifications that need to be met with respect to functionality, safety, and efficacy. Based on expert know-how, the Critical Quality Attributes (CQAs) are then identified and ranked for significance [27] (Figure 1, 1 – Identification of CQAs). This is followed by the identification of the Critical Process Parameters (CPPs) and Critical Material Attributes (CMAs). CPPs and CMAs are decided based on the impact they have on the chosen CQAs. In a traditional QbD approach, this impact is quantified through a series of wet-lab experiments. In this case (Figure 1, 2 – Identification of CPPs & CMAs) mathematical models that describe the process at hand and associated interactions can be used to study the system behavior and monitor the CQA-CPP-CMA cross-interactions. The same models can then serve as an *in-silico* testing platform to guide virtual experiments for the identification of good candidate condition sets, the combination of which will form the process DSp (Figure 1, 3 – Design Space).

Once the design (unit operations, materials) and operating conditions (process parameters) of the process have been identified, a control strategy that will maintain the process within the acceptable ranges needs to be developed (Figure 1, 4 – Process Control). The latter can also be supported by digital innovation, through the development of advanced control strategies that consider CQAs as target outputs, manipulating the ranges of CPPs. This process is iterative, and manufacturers are encouraged to improve the originally chosen conditions as they gain knowledge of the product and process (Figure 1, 5 – Continuous Process Improvement). In this work, we focus on the development of a model-based approach to support accelerated identification of the Design Space (Figure 1, 3-Design Space).

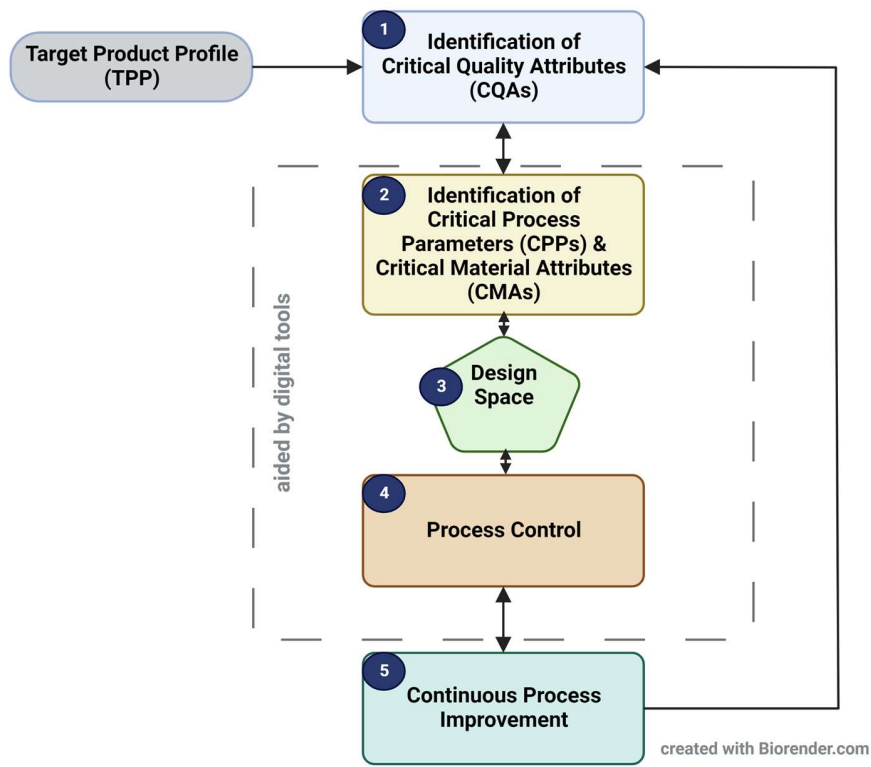

Figure 1. Framework for Quality by Digital Design (QbDD), whereby, the identification of Critical Process Parameters (CPPs), Critical Material Attributes (CMAs), Design Space, and Process Control are aided by computer-based (digital) tools.

**Process Systems Engineering in QbDD-driven approaches: Opportunities and Challenges**

The Process Systems Engineering (PSE) community has been developing tools and methodologies to enable model-based approaches that can accelerate the workflow as presented in Figure 1. Previous and current research spans various modeling techniques, from high-fidelity mechanistic modelling to machine learning (ML) and approaches, including adaptive sampling, constrained sensitivity analysis and stochastic approaches. Beyond (bio-)pharmaceutical applications, extensive research has been carried out focusing on both analytical [28] and probabilistic [29] approaches for the identification of the design space, as well as flexibility and optimization. An indicative list of PSE contributions with applications to (bio-) pharmaceuticals is presented in Table 1.

Table 1. Indicative list of PSE tools in (bio-) pharmaceutical design space identification.

| Approach | Reference |
|---|---|
| Probabilistic design space | Kusumo, et al. [30], Close, et al. [31], Close, et al. [32] |
| Design space identification via adaptive sampling | Rogers and Ierapetritou [33], Zhao, et al. [34] |
| Variance-based methods for design space identification | Kotidis, et al. [25], Nie, et al. [35], Vogg, et al. [36] |
| Machine learning-based design space identification | Rogers and Ierapetritou [37], Metta, et al. [38], Ding and Ierapetritou [39], Yang and Ierapetritou [40] |
| Design space identification following optimization approaches | Zhao, et al. [41], Bano, et al. [42] |

Despite these seminal contributions, there remain open challenges often specific to the modeling technique in use. In the case of mechanistic models, researchers are challenged by model complexity and computationally expensive formulations for data generation. On the other hand, fully data-driven approaches, although computationally cheaper, are challenged by limited transferability across products and critical materials (e.g., resins, filters etc.). There is therefore an opportunity to combine these modeling techniques to expedite computation while retaining process knowledge and transferability. To address this, we focus on integrating high-fidelity process modeling with ML for improved data generation and augment the approach by introducing computational geometry to explicitly identify the DSp and its boundaries.

## 2  A model-based approach for Design Space identification

We present a framework to assist QbDD and specifically the identification of the process DSp (Figure 1, 3 – Design Space). We propose a model-based approach, whereby process models are used for computational experiments to monitor the behavior of the chosen unit operation under a range of process conditions. The framework (Figure 2), employs a set of model-based tools to generate, classify and quantify sets of design and operating conditions that can yield within-spec products and processes. It can help accelerate process development by quantitatively linking operating variables, design parameters, and/or measured system disturbances in a single problem, to KPIs. Specifically, it comprises three main steps: (1) Model Development and Problem Formulation, (2) DSp Identification, and (3) DSp Analysis. Three options for the identification step are proposed for different stages of process design depending on the availability of prior knowledge (Figure 2 – 2. a, b, c). The mathematical details are presented in this section. The supporting computational package is accessible as open source through a user-friendly Python package (*dside*) published on PyPI and detailed at https://github.com/stvsach/dside.

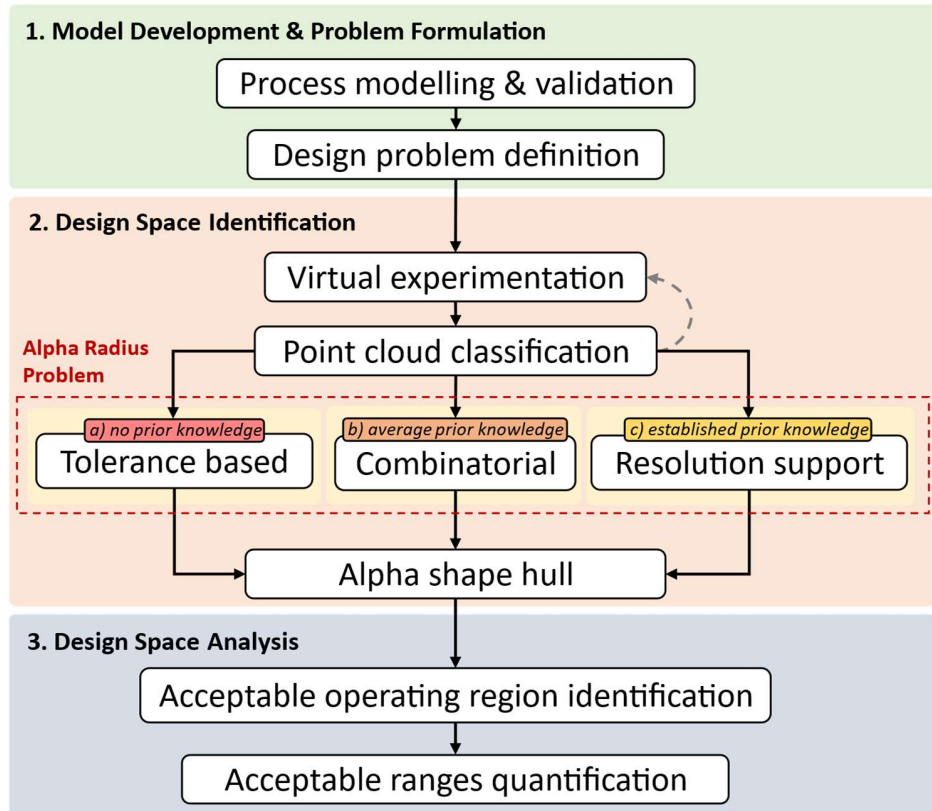

Figure 2. Schematic of the design space analysis framework.

The main outputs of the framework are recorded in the form of lookup maps, constructed by point clouds. Figure 3 depicts a simplified, 2D graphical illustration of the outputs that can be generated from this framework. Based on the constraints applied for each of the generated maps, we define the following terminology:

- *Knowledge Space (KSp)* (Figure 3, orange-shaded area) as the entire dataset (Figure 3, grey points) generated by varying the Design Decisions (DDs) within their feasible operating bounds (Figure 3, orange lines) as defined by the process and recording the values of the KPIs of interest.
- *Design Space* (Figure 3, green-shaded area) as a sub-region within the KSp defined by a set of points that satisfy process performance constraints imposed on the KPIs. The green lines (Figure 3) are input constraints imposed on the system to ensure that KPIs remain within operator-defined specifications.

- *Acceptable Operating Region (AOR)* (Figure 3, red-shaded area) as a region surrounding a nominal operating point (NOP) (Figure 3, purple point) of interest which guarantees operation inside the DSp.

- *Multivariate Proven Acceptable Range (MPAR)* as the acceptable variation range of an input for which operation is guaranteed to remain within spec (Figure 3, red lines).

It should be noted that Figure 3 is a simplification of the outputs generated as those are often n-dimensional maps depicted in 2D or 3D graphs as explained below.

In addition, computer-based models can be embedded in a QbD approach, to enhance product and process knowledge designing tailored experiments on the CQAs and CPPs of high interest. This translates into further accelerating the process development process and supporting risk-based decisions on experimental activities that need to be prioritized. Such models can be used to quantitatively process measurement data and be used as cost-efficient experimental platforms. In this context, previously published works report up to a 7-fold reduction in the absolute number of experiments that need to be carried out for the identification of the DSp [14].

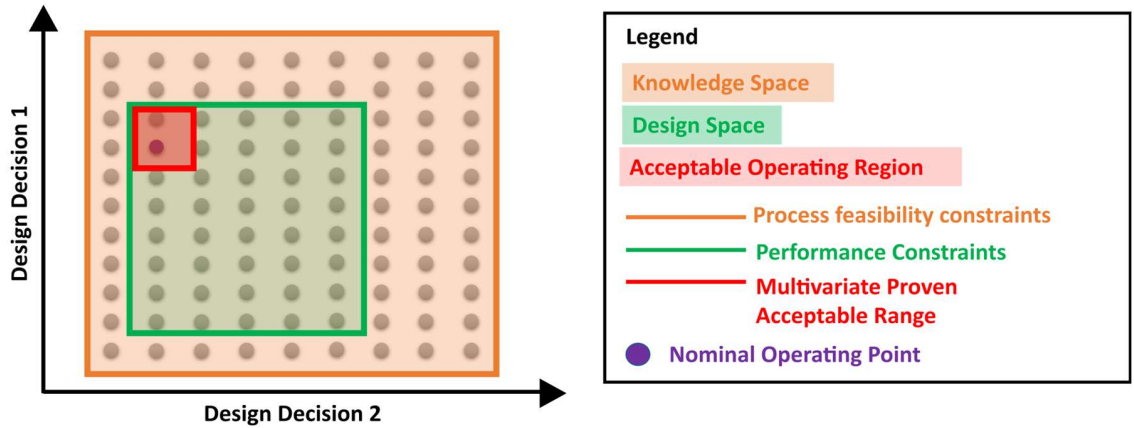

Figure 3. Graphical visualization of the lookup maps generated as framework outputs. The orange-shaded area corresponds to the KSp that is generated by varying the Design Decisions within their feasible range and monitoring the KPIs. The KSp boundaries correspond to the feasible bounds of the Design Decisions. The green-shaded area illustrates the DSp which is a collection of points satisfying performance constraints imposed on the KPIs. The green lines are Design Decision constraints imposed on the system to ensure that KPIs remain within operator-defined specifications. The red-shaded areas depict the Acceptable Operating Region (AOR) which is defined as a region surrounding a nominal operating point (NOP) (purple point) of interest that guarantees operation inside the DSp. The boundaries (red lines) depict the acceptable variation range of an input for which operation is guaranteed to remain within spec (Multivariate Proven Acceptable Range (MPAR)).

In this section (2.1-2.3) we provide detailed information on the mathematical problem formulations used in the framework. This is followed by the demonstration of the framework capabilities on an industrially relevant case study of a chromatographic separation unit and overall conclusions (Section 3).

## 2.1 Step 1. Model Development and Problem Formulation

### 2.1.1 Process Modelling and Validation

In this step, the mathematical model of the process is formulated, parameterized, and validated using experimental data. The model can be mechanistic, hybrid, or fully data driven [13, 43-45]. This step can be performed in any suitable platform, such as MATLAB, Python, or gPROMS ModelBuilder®.

### 2.1.2 Problem Definition

Here, the master design problem is identified (Equation (1)). The Design Decisions (DDs) and the KPIs of interest are chosen. Feasible bounds for the respective categories are selected. DDs can be operating variables, design parameters, or measured system disturbances.

$$\boldsymbol{y} = f(\boldsymbol{\theta})$$
$$\boldsymbol{\theta}_L \leq \boldsymbol{\theta} \leq \boldsymbol{\theta}_U \tag{1}$$
$$\boldsymbol{g}(\boldsymbol{y}) \leq 0$$

where $\boldsymbol{\theta} = [\theta_1, \theta_2, \ldots \theta_{n_\theta}]$ is the vector of design decisions of size $n_\theta$, $\boldsymbol{\theta}_L$ and $\boldsymbol{\theta}_U$ are the lower and upper bounds of the design decisions respectively chosen based on the feasibility constraints, $\boldsymbol{y} = [y_1, y_2, \ldots, y_{n_y}]$ is the vector of monitored KPIs of size $n_y$, $f$ is the process model, and $\boldsymbol{g} = [g_1, g_2, \ldots, g_{n_g}]$ is the vector of performance constraints with respect to the KPIs.

## 2.2  Step 2. Design Space Identification

The problem formulated in Step 1 is used as a basis for the following steps.

### 2.2.1  Virtual experimentation for Knowledge Space generation

The validated process model is used to generate the required dataset via the Sobol sequence within the space as defined in Step 1 [46]. The use of the Sobol sequence can allow full parallelization of the computational burden. Equation (2) shows the input generation strategy employed during the sampling procedure.

$$\begin{aligned}\theta_{in} &= Sobol(dim, \boldsymbol{\theta}_L, \boldsymbol{\theta}_U, sp) \\ &= \{\boldsymbol{\theta}_i : i = 1,2,3, \dots, n_p\}, \quad n_p = 2^{sp}\end{aligned} \quad (2)$$

where $\theta_{in}$ is the set of inputs, $\boldsymbol{\theta}_i$, generated based on the Sobol sequence [46]. The Sobol sequence is an n-dimensional quasi-random sequence that samples the search space uniformly with low discrepancy, while the sequence generation is based on the power of 2. The sequence can be tailored to $dim$-dimensions based on the number of design decisions ($n_\theta$) and scaled with respect to the lower and upper bounds of the design decisions ($\boldsymbol{\theta}_L$ and $\boldsymbol{\theta}_U$). Here, $sp$, determines the number of inputs generated ($n_p$). The generated inputs ($\theta_{in}$) are used for virtual experimentation to obtain the KSp dataset.

### 2.2.2  Point cloud classification and Bounds Refinement

The KSp (Step 2.1) is segmented into a feasible and an infeasible point cloud region, defined as follows (Equation (3)).

$$P_{KSp} = \{\boldsymbol{\theta}_i \in \theta_{\text{in}}: i = 1,2,3, \dots, n_p\}$$

$$P_{sat} = \{\boldsymbol{\theta}_i \in P_{KSp}: \boldsymbol{g}(\boldsymbol{y}(\boldsymbol{\theta}_i)) \leq 0\} \quad (3)$$

$$P_{vio} = \{\boldsymbol{\theta}_i \in P_{KSp}: \boldsymbol{\theta}_i \notin P_{sat}\}$$

$P_{KSp}$ is the KSp point cloud. The satisfied point cloud, $P_{sat}$, contains a subset of the KSp points satisfying all performance constraints, while the violated point cloud, $P_{vio}$, contains the remaining points that do not satisfy one or more of the performance constraints in $\boldsymbol{g}$. One can think of the data points in the point clouds as Cartesian coordinates that can be projected on a Euclidean space (Figure 4).

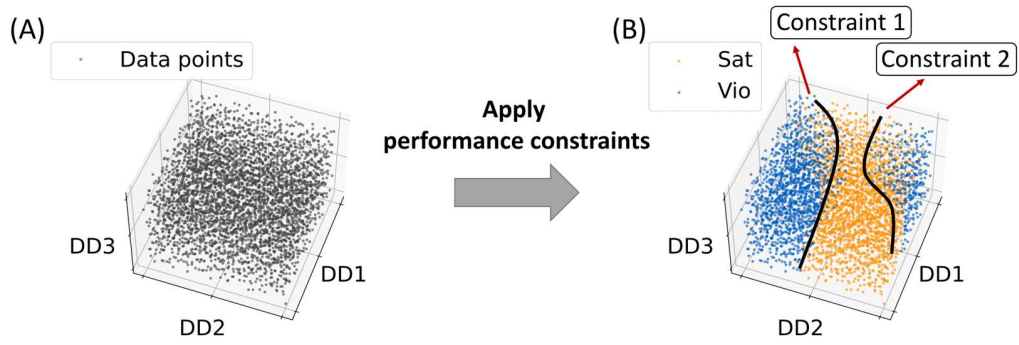

Figure 4. Example problem screened with the highlighted arbitrary non-linear performance constraints to obtain the point cloud of interest. (A): Black points show the data points obtained after the virtual experimentation that form the KSp, (B): the satisfied point cloud ($P_{sat}$) is shown with orange, and the violated point cloud ($P_{vio}$) is illustrated with blue points. The black lines correspond to Design Decision constraints imposed on the system to ensure that KPIs remain within specifications.

The grey point cloud ($P_{KSp}$) (Figure 4 (A)) forms the KSp that after the performance constraints are imposed, is segmented in the satisfied point cloud ($P_{sat}$) (Figure 4 (B), orange points) and the violated point cloud ($P_{vio}$) (Figure 4 (B), blue points). It should be noted that the number of $P_{vio}$ point clouds cannot be pre-identified and is dependent on the nature of the problem and the imposed constraints. At this stage, we can obtain graphical information regarding the feasibility of the space. This information can be particularly useful in the case of highly non-linear problems, where denser sampling may be required in specific regions. Steps 2.1 and 2.2 can be repeated to ensure the area of interest is well characterized (Figure 2, Step 2, dashed grey arrow).

### 2.2.3 Design Space Boundary

To enable deterministic quantification and analysis of the process performance metrics, a mathematical representation of the DSp boundaries (Figure 4 (B), continuous black lines) is needed. For this purpose, we employ computational geometry to identify and quantify the boundaries. The convex hull is a well-established technique typically used to generalize bounding polygons containing a set of points. However, convex hulls often lead to significant over/under-estimations when dealing with non-convex problems. Alpha shape (α-shape) is a generalization of a convex hull which can form bounding polygons that can accommodate non-convexity well. As most problems encountered in Chemical Engineering are often highly nonlinear, α-shapes are the most suitable way forward and, thus, considered in this paper. Both the convex hull and the α-shapes algorithms are well established [47-51] but for completion, we provide those in Appendix A.

### 2.2.4 The alpha radius Problem and the determination of a suitable α-shape

The objective of this step is to define a single, unified, suitable *α-shape* to be used as the DSp. We define a "*suitable α-shape*" as the α-shape that describes the largest geometrical space the points within which satisfy all imposed constraints. Often, in nonlinear problems, the identification of a single α-shape is challenging, primarily due to unexplored areas that result in disjointed spaces. One way to tackle this is to identify an appropriate alpha radius ($\alpha_r$). The alpha radius influences the accuracy with which an α-shape is formed and, in this case, whether this shape is a valid representation of the DSp [51]. The criticality of the alpha radius problem, its relationship with the problem nonlinearity, and its impact on the identification of the DSp are discussed in detail in Appendix A. Briefly, a low $\alpha_r$ value is likely to lead to an α-shape that contains no violations but is formed by >1 regions while increasing values of $\alpha_r$ would increase the probability of obtaining a single region shape containing points that violate one or more of the performance constraints. One may wish to promote the design of a unified α-shape by generating additional points around areas of high interest at the expense of higher computational cost.

To address this and integrate both avenues, in this framework, we propose three different methodologies that cater to different levels of detail and accuracy in the resulting DSp. These are the tolerance-based, the resolution support, and the combinatorial method. The key differences across these three methods lie in (1) the percentage of violated points allowed in the DSp and (2) the amount of data points required for the identification of an α-shape representative of the DSp. In the presented framework, the choice of a suitable alpha radius has been integrated with the choice of the methodology through Algorithm A2. The procedure is iterative and should be repeated until a suitable α-shape has been identified (Figure 2, Step 2, dashed red box). Below we summarize the key features of each method, the performance of which is assessed through the case study presented in Section 3 of the manuscript.

*2.2.4.1 Tolerance-based method*

This methodology is proposed for a swift analysis, as it allows for high alpha radius values to be chosen, requiring less computational effort. A possible downside of this is that the produced α-shape may contain points that violate one or more of the constraints. To implement this method, Algorithm A2 (Appendix A) can be adapted by setting a tolerance value ($v_{max\%}$ in Algorithm A2, Appendix A). This defines the percentage of violations allowed in the α-shape. Post-analysis would be required to assess whether the violations present in the α-shape are acceptable, for example when lying within the analytical error of monitoring equipment in use. A specific example of this and a discussion on what may be an acceptable violation is demonstrated through the case study in Section 3.

*2.2.4.2   Resolution support method*

A pathway to circumvent the challenges of disjointed spaces and/or violations within the final α-shape is the generation of additional points via computational experiments. Given that in most cases the high-fidelity models are highly complex, this method may result in higher computational cost. To decrease the computational effort required we train and use an Artificial Neural Network (ANN) as a data interpolator. The ANN is trained and tested on the first dataset generated in Step 2.1. Specifically, the dataset is separated into a training and a testing subset, where the former serves for the construction of the ANN, while the latter is used for validation. Following Steps 2.1 and 2.2, above, the validated ANN is then used for data generation ($P_{sat,extra}$) to enrich the information vector within the KSp ($P_{KSp}$).

The choice of the number of extra points required to be generated ($P_{sat,extra}$) is critical, as it defines the computational effort required, and can directly impact the quality of the generated α-shape, as discussed above. The generation of extra points can be carried out using the ANN following the inputs generated from Sobol sequence. As detailed in Section 2.2.1, the inputs are generated based on $2^n$, where $n \in (1, +\infty)$. To inform the choice of a suitable value for $n$, we run a scenario-based analysis, starting from $2^{10}$ simulations to obtain the enhanced dataset ($P_{KSp} + P_{sat,extra}$). If the number of resulting regions, as suggested by Algorithm A2 (Appendix A) is equal to one, then the search is concluded, otherwise, the procedure is repeated increasing the value of $n$, until a unified α-shape is formed. The accuracy of the ANN is important at this stage, as inaccuracies would likely introduce non-convexities, possibly leading to larger datasets required. For problems described by linear relationships, the ANN can be substituted by linear interpolation.

*2.2.4.3  Combinatorial method*

To exploit the advantages offered by the two previously presented methods, here, we propose their integration. For this, we follow the procedure described in the resolution support (Section 2.2.4.3) and adapt Algorithm A2 (Appendix A) to consider a value of $v_{max\%} > 0$. By relaxing the $v_{max\%}$ constraint, a unified α-shape can be formed using fewer points, hence decreasing the size of the $P_{sat,extra}$ generated by the ANN. This can enhance the computational speed toward the identification of a suitable α-shape and therefore the DSp. Nevertheless, similar to the tolerance method, the generated space would likely contain points that do not satisfy all the constraints. Post-processing would be required to assess the violations present in the α-shape.

## 2.3  Step 3. Design Space Analysis

Step 3 uses the DSp identified in Step 2 to assess process performance under different conditions and quantify its capability to handle variations in the assumed Design Decisions. Here, we define two metrics to streamline the analysis of the DSp:

1. The size of the identified region. This can be quantified by calculating the volumes of the tetrahedrons (3D α-shapes) or the area of triangles (2D α-shapes). The greater the value, the higher the flexibility of the given process.
2. The Multivariate Proven Acceptable Range (MPAR). MPAR is specific to a chosen Nominal Operating Point (NOP) of interest and quantifies the proven acceptable ranges of the Design Decisions such that the process operation remains inside of the DSp and within spec. Equation (4) shows the mathematical representation of the MPAR.

$$\mathbf{MPAR}_L \leq \boldsymbol{\theta}_{\text{NOP}} \leq \mathbf{MPAR}_U \quad \text{s.t.} \quad g(\boldsymbol{\theta}, \mathbf{y}(\boldsymbol{\theta})) \leq 0 \tag{4}$$

where $\mathbf{MPAR}_L$ and $\mathbf{MPAR}_U$ are vectors of length $n_\theta$ containing the MPAR of each design decision with respect to the NOP denoted by the vector of design decisions $\boldsymbol{\theta}_{NOP}$. To quantify this metric, a feasibility test is performed to ensure the NOP lies within the DSp. A bisection search is applied starting from the NOP to form a uniform polygon region (square for 2D and cube for 3D). This is expanded outwards and uniformly towards all directions until one of the vertices hits the DSp boundary (Appendix A, Algorithm A3). The resulting region is the acceptable operating region (AOR). Figure 5 illustrates a simplified, graphical representation, where the NOP is represented by a solid cross, while the acceptable operating region is shown by the cube with solid black lines. Operation within this AOR is expected to satisfy all constraints.

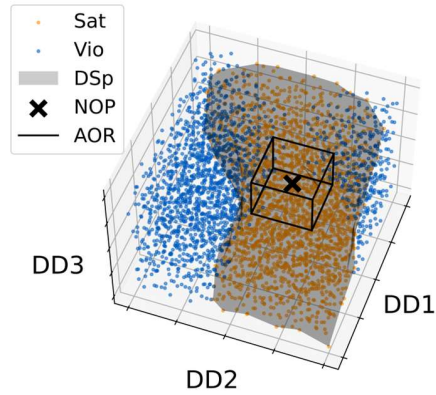

Figure 5. Example DSp with arbitrary design decisions (DD1, DD2 & DD3), KPIs, constraints, and nominal operating point (NOP). "Sat" are the satisfied computational samples while "Vio" are the violated samples. AOR: acceptable operating region with respect to the NOP.

Operation within the MPAR of each design decision guarantees satisfaction of all the constraints. In cases where there are more than one Design Decisions (e.g., Figure 5), this becomes a multivariate analysis and all design decisions can be varied simultaneously within their respective MPARs, guaranteeing feasibility for all possible combinations. Further analysis, such as the univariate Proven Acceptable Range (PAR [12]), focusing on examining the acceptable range of one of the design decisions while keeping the others constant can also be performed.

## 3 Use case: Protein A Chromatography

The capabilities of the framework described are assessed through its application to a Protein A chromatography process. Protein A chromatography (capture) is typically the first step in monoclonal antibody (mAb) downstream purification, used to process the harvest cell culture fluid obtained from the upstream bioreactor, containing the product and impurities. This step is used to capture the product removing most of the bulk protein impurities and reducing the sample volume, while its performance is assessed based on the resulting process yield.

A challenge often encountered in the design of biopharmaceutical downstream separation processes is the variability in the input mixture composition. This is associated with the cell-based upstream production system [5] and cannot be controlled. Therefore, there is an eminent need to design separation units that can handle feed variability. In this work, we follow the above-presented framework to identify the process DSp and quantify the capability of the process to handle feedstock disturbances.

### 3.1 Step 1. Model Development and Problem Formulation

**Process Modelling and Validation.** We consider the capture process presented by Steinebach, et al. [52] (Figure 6). This is a multicolumn, cyclic process considering three main steps for continuous processing of the feed. At first (Step A), both columns are interconnected, with column 1 being fed with the fresh feed while column 2 receives the breakthrough of column 1. In Step B, column 1 is being washed, while column 2 is fed with the breakthrough of column 1 and fresh feed at the same time. In the last step (Step C), the columns are operated in batch, where column 1 is eluted and regenerated while column 2 is fed with fresh feed. Then the columns switch positions, and the sequence is repeated.

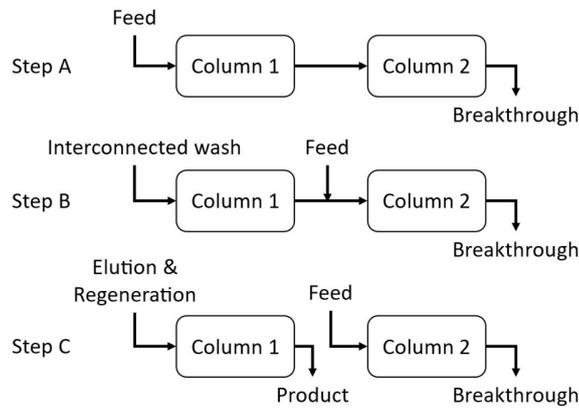

Figure 6. Multicolumn Protein A Chromatography process schematic adapted from Steinebach, et al. [52].

The process is described by a high-fidelity mechanistic model, considering lumped mass transfer kinetics with dual site Langmuir adsorption isotherm [52]. The model consists of Partial Differential Equations (PDAEs) that, after spatial axis discretization using 50 collocation points, result in a total of 808 Ordinary Differential Equations (ODEs) and 812 algebraic equations [52]. Details on the model equations are presented in Appendix B. The process and model have been validated for mAb concentrations of 0.2 – 0.77 mg/mL and feed flow rates of 0.5 – 1.5 mL/min [52]. Here we implement the model in Python 3.9.9, using central finite difference for the discretization of the column length and we solve the ODEs using SciPy 1.8.0.

**Problem Definition.** In this study, the objective is to investigate the flexibility and performance of the process under variable feed concentration. The variables considered are mAb concentration in the feed stream ($c_{feed}$) [disturbance], feed volumetric flow rate ($Q_{feed}$) [operational variable], and the column switching time ($T_{switch}$) [design variable]. The combination of design decisions and disturbances allows for simultaneous investigation of the performance of a design of the process and the operational flexibility of the corresponding design. The KPIs considered in the study are yield and productivity. The feasibility constraints are chosen with respect to the validated fidelity of the model, while we impose a yield constraint ≥ 99% and a productivity constraint ≥ 4 mg ml$^{-1}$ h$^{-1}$. Assuming limited prior knowledge, we choose wide ranges for the bounds of the Design Decisions. A small number of samples is generated for preliminary exploration (Table 2).

## 3.2 Step 2. Design Space Identification

**Initial Bounds Exploration.** Using the Sobol sequence, a total of 512 virtual experiments are performed to collect the initial dataset (Figure 7 (A)).

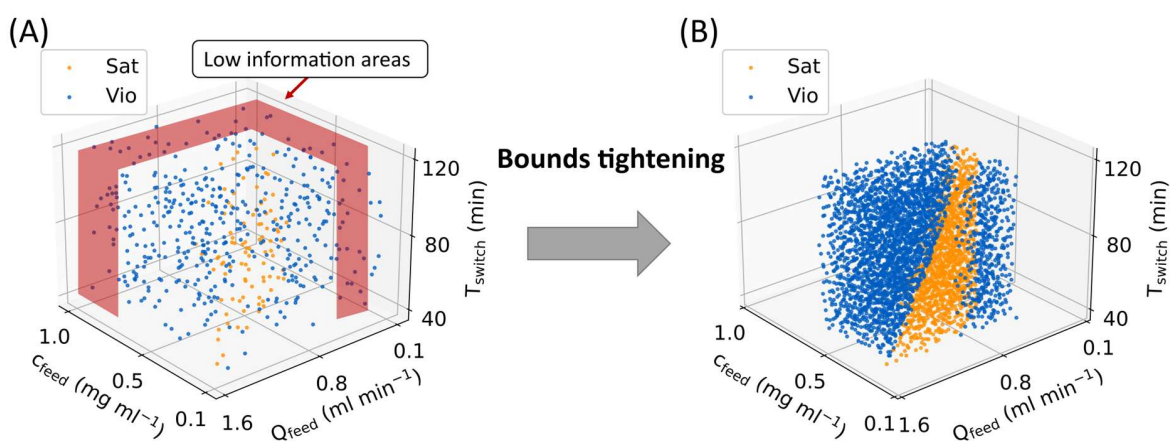

Figure 7. (A) Initial sampling performed to gather 512 points for surveying the knowledge space, (B) Bounds are then tightened to focus on the area of interest with more satisfied points.

It can be observed (Figure 7 (A)) that there are areas with low density of satisfied points, particularly around $c_{feed}$ > 0.7 mg ml$^{-1}$ and $Q_{feed}$ < 0.5 ml min$^{-1}$, indicating that operating beyond these values, the present process setup will likely yield out-of-spec products. To intensify the search within the areas of high interest that are the ones that satisfy the yield and productivity constraints, the feed concentration and flow rate bounds are tightened, and the number of sampling points is increased (Table 2).

Table 2. Initial and refined design decision bounds.

| Design Decision | Lower Bound | Upper Bound |
|---|---|---|
| Initial Bounds | | |
| $c_{feed}$ (mg ml-1) | 0.10 | 1.00 |
| $Q_{feed}$ (ml min-1) | 0.10 | 1.57 |
| $T_{switch}$ (min) | 40 | 120 |
| Refined Bounds | | |
| $c_{feed}$ (mg ml-1) | 0.21 | 0.63 |
| $Q_{feed}$ (ml min-1) | 0.50 | 1.50 |
| $T_{switch}$ (min) | 40 | 120 |

**Point cloud classification.** The refined bounds (Table 2) are used to sample 4096 data points which will be used as the main dataset for this study. Following the application of the yield and productivity constraints, the point clouds (Figure 7 (B)) indicate a continuous region exists that includes only the samples that satisfy all constraints. Next, we assess and compare the capabilities of the three methodologies as presented in Sections 2.2.4.1 – 2.2.4.3.

### 3.2.1 Tolerance-Based Method

A tolerance value for the number of violations is imposed to acquire a unified α- shape. Grid search is implemented to find the appropriate tolerance value to be used. The grid starts with $v_{max\%}$ = 0 and is incrementally increased by 0.25% until the alpha shape formed contains only one single region. Figure 8 (A) – (D) illustrates the results of the identified alpha shapes via the iterative grid search. Identifying the DSp with zero tolerance results in 13 disjointed regions (Figure 8 (A)). By increasing the violation tolerance ($v_{max\%}$, Algorithm A2, Appendix A), the algorithm is allowed to use larger values of alpha radius, leading to a reduced number of regions (Figure 8 (B) to (D)).

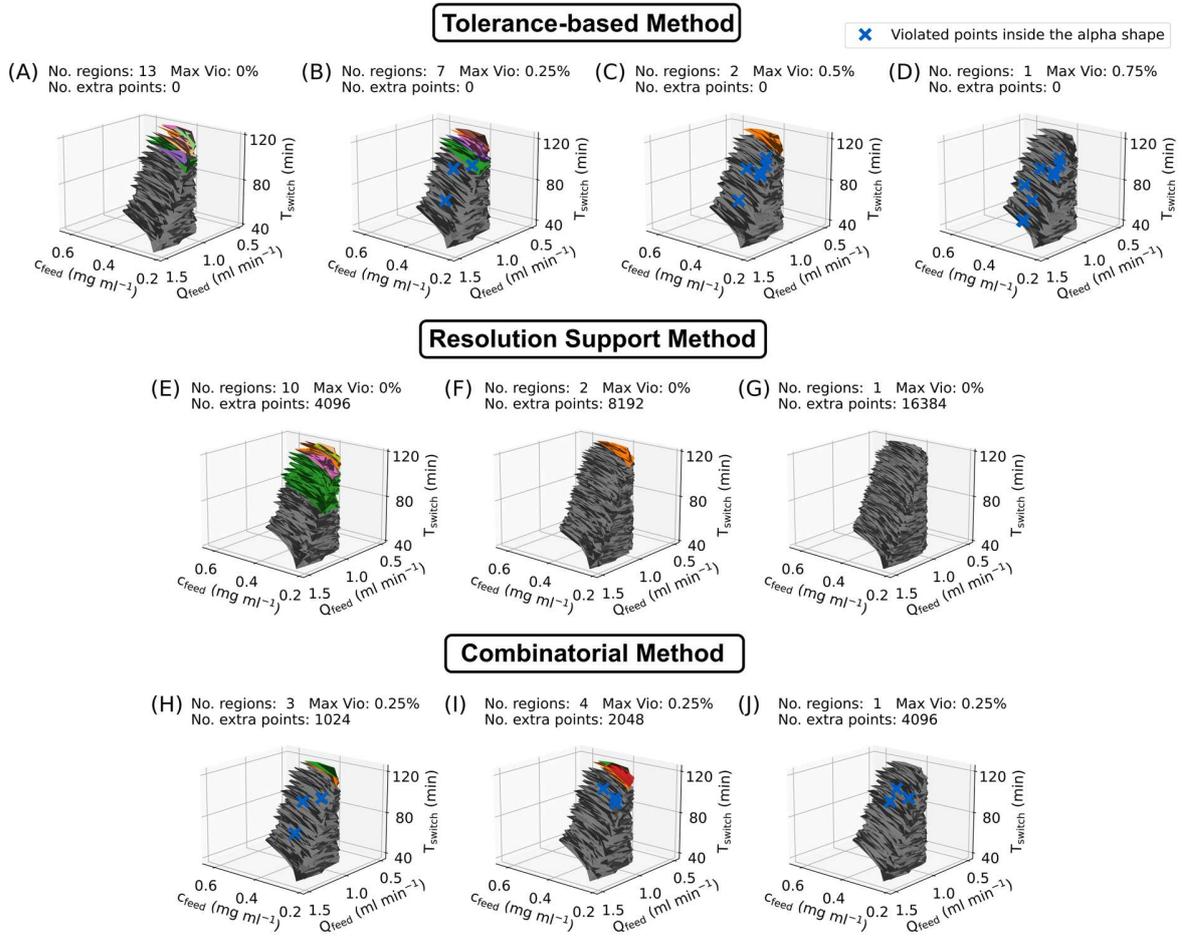

Figure 8. Alpha shapes defined with different methods. (A) – (D) shows the tolerance-based method at different tolerance values ($v_{max\%}$): (A) zero tolerance, (B) 0.25% resulting in 3 violated points, (C) 0.5% resulting in 6 violated points, and (D) 0.75% resulting in 9 violated points. (E) – (G) is the resolution support method with increasing numbers of extra points: (E) $2^{12}$ = 4096 extra points, (F) $2^{13}$ = 8192, and (G) $2^{14}$ = 16384. The combinatorial method is shown in (H) – (J) at a tolerance of 0.25% with (H) $2^{10}$ = 1024 extra points, (I) $2^{11}$ = 2048, and (J) $2^{12}$ = 4096. Each color represents a disconnected region.

Before the formed alpha shape can be declared as a suitable representation of the DSp, an assessment of the nature and level of the violations present in the shape should be carried out. Table 3 shows in detail the nine violated points lying inside of the a-shape, along with their corresponding KPI values.

Table 3. Details on the violated points lying inside of the alpha shape with 0.75% tolerance (Figure 8(D)) in the maximum percentage of violation points inside the shape.

| No | $c_{feed}$ (mg ml$^{-1}$) | $Q_{feed}$ (ml min$^{-1}$) | $T_{switch}$ (min) | Yield (%) | Productivity (mg ml$^{-1}$ h$^{-1}$) |
|---|---|---|---|---|---|
| 1 | 0.33 | 0.88 | 104.5 | 98.98 | 5.37 |
| 2 | 0.52 | 0.83 | 67.7 | 98.91 | 7.87 |
| 3 | 0.42 | 1.12 | 45.0 | 98.99 | 8.56 |
| 4 | 0.34 | 0.94 | 91.3 | 98.95 | 5.87 |
| 5 | 0.32 | 0.93 | 99.7 | 98.99 | 5.41 |
| 6 | 0.32 | 0.99 | 90.3 | 98.91 | 5.81 |
| 7 | 0.45 | 0.81 | 87.1 | 98.86 | 6.64 |
| 8 | 0.43 | 1.00 | 61.6 | 98.93 | 7.67 |
| 9 | 0.43 | 1.11 | 44.8 | 98.96 | 8.69 |

All nine points satisfy the productivity constraint but failed to satisfy the yield threshold (Table 3). Nevertheless, the recorded violation in the yield constraint can be considered negligible in all cases (≤0.14%), as it lies within the measurement error typically reported in process analytical tools (PATs) used in biopharmaceutical process monitoring (~1% [53]). Therefore, the α-shape formed in Figure 8 (D) can be considered a valid representation of the DSp.

### 3.2.2 Resolution Support

In this case, an Artificial Neural Network is trained to act as a non-linear interpolator to increase the resolution of the dataset. For this case study, three hidden layers each of size 256 are used, with the rectified linear unit (ReLU) activation function. The ANN has three inputs, namely: the design decisions ($c_{feed}$ – feed mAb concentration, $Q_{feed}$ – feed mAb flowrate, and $T_{switch}$ – column switching time) and uses the KPIs (yield and productivity) as outputs. Input and output min-max normalization is used to increase the fitting performance for the ANN. The neural network was trained using the Adam optimizer with mean square error (MSE) as the loss function for 50,000 epochs and a learning rate of $5 \times 10^{-6}$. The training is performed in Python using the Pytorch library with an NVIDIA GeForce RTX™ 3070 Ti GPU, with a CPU time of 12 minutes. It is worth noting that in this case, the CPU time required for a simulation of a single data point using the high-fidelity model is 4 minutes, which is 1/3 of the time required to train the ANN. The resulting performance is shown in Figure 9.

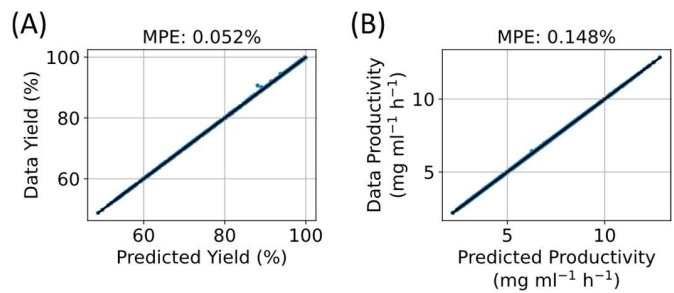

Figure 9. Trained neural network performance (A) yield prediction, (B) productivity prediction with their mean percentage error (MPE) shown.

As illustrated in Figure 9 (A) and (B), the ANN prediction performance is satisfactory for both KPIs of interest. The mean percentage error (MPE) against the testing dataset for the yield prediction is 0.052% and for the productivity is only 0.148% and can be considered satisfactory.

Next, we implement a grid search to identify the appropriate number of extra data points required for a smooth point cloud. Starting from $2^{10}$ extra points generated based on inputs from the Sobol sequence, the power is increased by one until an alpha shape with just one region is formed. Figure 8 (E) – (G) shows the last three iterations of the alpha shapes formed with their regions visualized with different colors. Figure 8 (G) illustrates the final alpha shape, whereby 16,384 additional points ($P_{sat,extra}$) were required to form exactly one single region without any violations.

### 3.2.3   Combinatorial Method

For the combinatorial method, a tolerance value of 0.25% is used, with no extra points, this results in 7 regions (Figure 8 (B)). A grid search is then used to identify the appropriate number of extra points to be used for the identification of a unified α-shape (Figure 8 (H) – (J)). All violations inside the shape for all cases (Figure 8 (H), (I), and (J)) is well within 1% of PAT error with the largest being 0.36% (data shown in Appendix C, Table C1). The combinatorial method obtains a valid α-shape with lower violations inside the shape and a smaller number of extra points in contrast to the previous two methods. As the alpha shape in Figure 8 (J) contains only a single region and the violations are less than PAT error, it is considered as a valid DSp.

### 3.2.4 Comparative Analysis

All three methodologies converge to a unified α-shape that can be used as a representation of the DSp of the process. In this example, a 3-D problem is presented, where the DSp axes are the feed concentration, feed flow rate, and the switching time, while the space is quantified in mg (Equation (5)).

$$c_{feed} \times Q_{feed} \times T_{switch} = \frac{mg}{ml} \times \frac{ml}{min} \times min = mg \tag{5}$$

Table 4 summarizes the quantified DSp across the three methods presented. It can be observed that each method results in spaces of different sizes. This can be translated into a dependence of the size on data availability. Specifically, the DSp sizes are equal to 5.80 mg, 6.97 mg, and 5.67 mg using the tolerance-based, resolution support and the combinatorial method, respectively. In the combinatorial method, the tolerance value has a greater impact on the DSp size as the quantified space is smaller than that defined by the tolerance method. On the contrary, the space designed with the resolution support method is greater in size compared to all other methods, with zero recorded violations. This highlights the importance of high data resolution and, despite the computational cost, the advantage of the resolution support method when process knowledge is available.

Table 4. Design space identification method comparison. Computation time measured was using an Intel i7 11700K CPU machine with 32 GB of RAM.

|  | Tolerance | Resolution Support | Combinatorial |
|---|---|---|---|
| Space size (mg) | 5.80 | 6.97 | 5.67 |
| Alpha radius (mg$^{1/3}$) | 2.49 | 0.62 | 0.97 |
| No. Violations in DSp (dimensionless) | 9 | 0 | 3 |
| No. Extra Points (dimensionless) | 0 | 16,384 | 4,096 |
| No. Satisfied Points (dimensionless) | 1,269 | 6,364 | 2,548 |
| DSp Computation time (mins) | 1.0 | 5.6 | 1.9 |
| Hyperparameter Analysis time (mins) | 3.5 | 12.8 | 4.5 |

Despite the benefit of a smoother, continuous space, the resolution support method relies on the generation of additional data points that may increase the computational effort required. For an objective computational cost comparison, in this analysis, the upper and lower bound, and bisection tolerance are the same across the three methods. Similarly, the starting number of extra points used in both resolution support and combinatorial method is the same (Table 4). As anticipated, the resolution support method takes the longest time to complete (Table 4), due to the higher number of iterations required to find a zero-violation DSp, as it processes a significantly larger dataset. It should be noted in the case of the resolution support, the time taken to complete the last iteration with $2^{14}$ extra points accounts for almost half of the time taken for the full analysis. It should be highlighted that in a real-work application of the presented approach, only one method should be chosen and followed, based on the level of prior knowledge and the desired accuracy.

## 3.3 Step 3. Design Space Analysis

The operational flexibility of the process is assessed using the DSp as identified in Step 2. An acceptable operating region (AOR) with respect to a combination of nominal values for the design decisions can be quantified. Here we choose an arbitrary nominal point of interest: $c_{feed} = 0.40\ mg \cdot ml^{-1}$, $Q_{feed} = 0.80\ ml \cdot min^{-1}$ and $T_{switch} = 70\ min$.

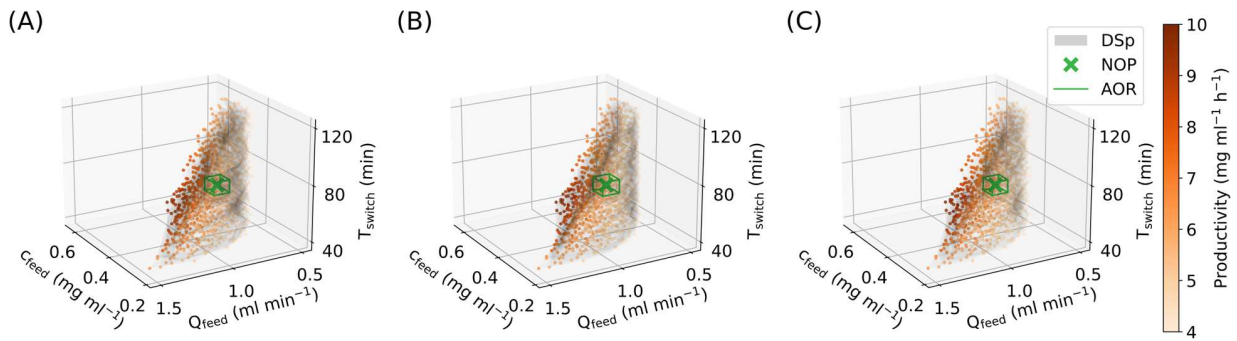

Figure 10 An acceptable operating region formed with respect to the nominal point using the three different methods: A) tolerance, B) resolution support, and C) combinatorial. The satisfaction points are colored with a productivity heat map. All points shown satisfied both yield and productivity constraints.

Figure 10 shows the AOR as identified using the different DSp formed with the three methods. The tolerance and combinatorial methods agree with respect to the identified AORs, while the resolution support method identifies a more optimistic scenario of a larger region. This difference can be attributed to the larger number of extra points resulting in a smoother boundary. By identifying the uniform AOR, the MPAR of each manipulated variable is also quantified, the boundaries of which correspond to the acceptable deviation in the feed composition ($c_{feed}$), feed flow rate ($Q_{feed}$) and switching time ($T_{switc}$ ).

Figure 11 and Table 5 illustrate the resulting flexibility metrics for each method followed in Step 2. Due to the different alpha radius values and different sets of points used in the three identification methods, the reported DSp and flexibility metrics are slightly different. Interestingly, the identified AOR resulting from the tolerance method is the same as the one defined using the combinatorial method. This can be attributed to the identified DSp being relatively similar in the number of tetrahedrons and space size. It is also evident that the number of tetrahedrons affects the CPU time required to do the analysis (Table 5). However, the combinatorial method still offers an advantage in the ability to increase the resolution within any region of the space. This enables the analysis of space specific KPIs which is detailed in the next subsection.

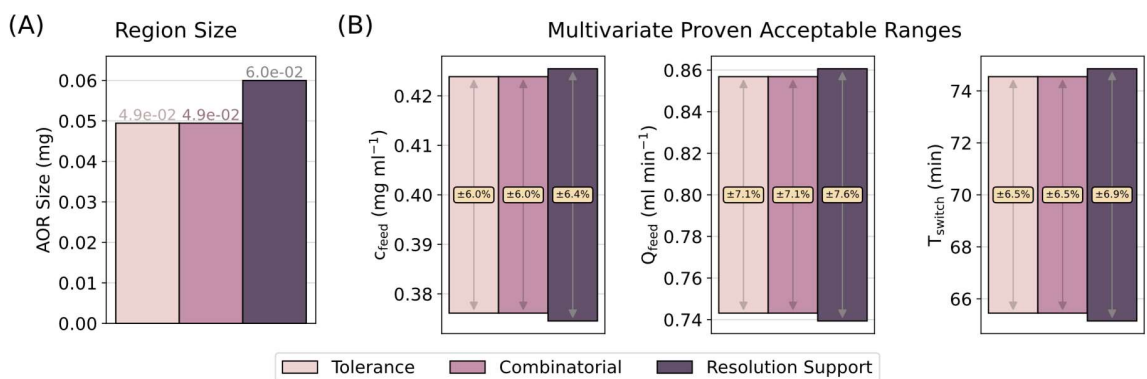

Figure 11. Method flexibility metrics comparison: A) Acceptable Operating Region (AOR) size and (B) Multivariate Proven Acceptable Ranges (MPAR) of $c_{feed}$, $Q_{feed}$, and $T_{switch}$, for the different methods.

Table 5. AOR and MPAR details based on the different methods.

|  | Tolerance | Combinatorial | Resolution Support |
|---|---|---|---|
| No. Tetrahedrons (dimensionless) | 4,737 | 9,831 | 29,022 |
| AOR Computation time (s) | 5.3 | 10.9 | 34.9 |
| AOR size (mg) | 0.049 | 0.049 | 0.060 |
| Range $c_{feed}$, Nominal: 0.4 mg ml$^{-1}$ | ±0.024 | ±0.024 | ±0.025 |
| Range $Q_{feed}$, Nominal: 0.8 mg min$^{-1}$ | ±0.057 | ±0.057 | ±0.061 |
| Range $T_{switch}$, Nominal: 70 min | ±4.5 | ±4.5 | ±4.9 |

### 3.3.1 Space Specific KPI Analysis

The presented framework allows for analysis of the KPIs inside any constrained space within the KSp. The distribution of the productivity inside the DSp and the acceptable operating region is illustrated in Figure 12, while Table 6 summarizes the associated KPI analysis metrics. For a fair comparison, in the case of the tolerance method, no neural network is trained and therefore the analysis is limited to the original data points obtained.

It can be observed in Figure 12 that, all methods agreed on the fact that a narrow distribution is reported for a target productivity of 8 mg ml$^{-1}$ h$^{-1}$, while it is broader in the case of lower productivity values (5.0 mg ml$^{-1}$ h$^{-1}$). In practice, this translates into fewer operating points that can achieve a productivity of 8 mg ml$^{-1}$ h$^{-1}$ and therefore potentially challenging process optimization. Although all methods have a similar average productivity (Table 6), the combinatorial and resolution support method allows for the exploration of a larger percentage of the available DSp and refine the analysis by ANN interpolation gaining more confidence around the flexibility of the process.

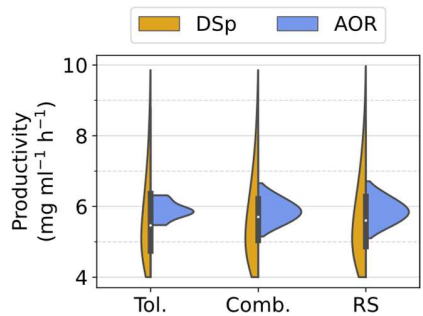

Figure 12. Productivity analysis in the Design Space and Acceptable Operating Region (AOR) for the Tolerance (Tol), Resolution Support (RS), and Combinatorial (Comb) method. The AOR designed here is based on the chosen nominal point $c_{feed} = 0.40\ mg \cdot ml^{-1}$, $Q_{feed} = 0.80\ ml \cdot min^{-1}$ and $T_{switch} = 70\ min$.

Table 6. Space specific Key Performance Indicator analysis details.

|  | Tolerance | Combinatorial | Resolution Support |
|---|---|---|---|
| DSp Maximum Productivity (mg ml$^{-1}$ h$^{-1}$) | 9.86 | 9.86 | 9.97 |
| DSp Average Productivity (mg ml$^{-1}$ h$^{-1}$) | 5.68 | 5.67 | 5.67 |
| DSp Minimum Productivity (mg ml$^{-1}$ h$^{-1}$) | 4.00 | 4.00 | 4.00 |
| No. samples in AOR (dimensionless) | 7 | 7 | 9 |
| No. samples in AOR after support (dimensionless) | N/A | 1031 | 1033 |
| AOR Maximum Productivity (mg ml$^{-1}$ h$^{-1}$) | 6.32 | 6.66 | 6.71 |
| AOR Average Productivity (mg ml$^{-1}$ h$^{-1}$) | 5.89 | 5.87 | 5.87 |
| AOR Minimum Productivity (mg ml$^{-1}$ h$^{-1}$) | 5.47 | 5.15 | 5.10 |

Overall, all three methods result in reliable DSp. Their key differences lie in (1) computational speed and (2) the percentage of violations present in the identified space. On occasions where the process at hand is novel and no prior knowledge is available, one may choose to employ the tolerance method as it enables a wide and swift search of the feasible space. In such cases, post-processing of the violations present in the DSp is required to assess whether those can be deemed acceptable. This depends on the analytical measurement error as well as the fidelity of the model used for the computational experiments. On the other end, in cases where prior knowledge is well-established and the system specifications have been narrowed down, the resolution support method is a better fit as it provides DSps of high resolution and fidelity. In such cases, the bounds of the Design Decisions are expected to be tighter and therefore the computational burden is expected to remain tractable. For cases that lie in the middle, where prior knowledge is available but not extensive, the combinatorial method is a suitable choice.

## 4 Towards a good candidate design

Typically, during process development, different operating points are chosen and assessed for feasibility and performance. Here, we demonstrate how the above-presented framework can assist a systematic comparison of different design and operating conditions in terms of performance as well as operating flexibility. We select a second, arbitrary operating point (NOP2) within the DSp obtained by the combinatorial method (Figure 10 (C)), the operating conditions of which are: $c_{feed} = 0.55\ mg \cdot ml^{-1}$, $Q_{feed} = 0.75\ ml \cdot min^{-1}$ and $T_{switch} = 50\ min$, with a smaller AOR (Figure 13 (A)). In this case, $AOR_{NOP2} < AOR_{NOP1}$ by 24%, translating into a less flexible process in case of operation under the conditions of NOP2 (Table 7). This flexibility is determined by the level of disturbances the process can handle in the design decisions, without being driven out-of-spec. In comparison to NOP1, an operating profile under the conditions of NOP2 would allow the process to handle 8.8% less variation in the operating values of $c_{feed}$, $Q_{feed}$, and $T_{switch}$, respectively (Figure 13 (B) and Table 7).

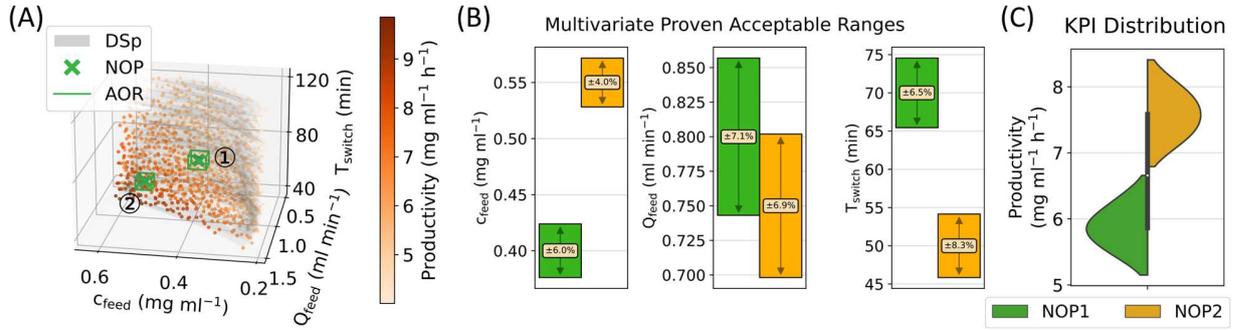

Figure 13. Investigation of a good candidate design. ①: NOP1 base case investigated in the previous section, ②: NOP2 good candidate design.

Table 7. Comparison of NOP1 & NOP2 (percentages refer to the change from NOP1 to NOP2).

|  | NOP1 | NOP2 |
| --- | --- | --- |
| AOR size (mg) | 0.049 | 0.038 (−24%) |
| MPAR: $c_{feed}$ (mg ml$^{-1}$) | 0.40 ± 0.024 | 0.55 ± 0.022 (−8.8%) |
| MPAR: $Q_{feed}$ (mg min$^{-1}$) | 0.80 ± 0.057 | 0.75 ± 0.052 (−8.8%) |
| MPAR: $T_{switch}$ (min) | 70 ± 4.5 | 50 ± 4.1 (−8.8%) |
| Maximum AOR Productivity (mg ml$^{-1}$ h$^{-1}$) | 6.66 | 8.41 (+26%) |
| Average AOR Productivity (mg ml$^{-1}$ h$^{-1}$) | 5.87 | 7.59 (+29%) |
| Minimum AOR Productivity (mg ml$^{-1}$ h$^{-1}$) | 5.15 | 6.79 (+32%) |

Figure 13 (C) illustrates the productivity distributions for both NOP1 and NOP2. It can be observed that NOP2 operation results in higher absolute productivity, which can be attributed to the higher feed concentration and lower switching time. In addition, the resulting productivity within $AOR_{NOP}$ has a wider range (±1.62 mg ml$^{-1}$ h$^{-1}$) of variation, compared to the one within $AOR_{NOP}$ (±1.51 mg ml$^{-1}$ h$^{-1}$) (Table 7), quantified in an average increase of 29% when operating at the conditions of NOP2.

One can conclude that NOP1 offers greater process flexibility and robustness when it comes to inherent disturbances (AOR size), while NOP2 would guarantee higher productivity targets. This is a trade-off commonly encountered in industrial case studies, especially during process development, where different sets of conditions are screened with little or no prior knowledge of the process performance. It is worth noting that the methodologies presented here can be viewed as customizable tools that should be used in a staged fashion.

# 5 Conclusions

We presented a model-based approach for the identification and analysis of process DSp that can be used to inform process development and guide the choice of suitable candidate process conditions. The framework, available via the Python package '*dside',* exploits the fully parallelizable Sobol sequence to create a data cloud (*knowledge space*) that describes the process performance with respect to chosen operating variables. Imposing feasibility and performance constraints, this can be tailored to provide a DSp containing sets of process conditions that satisfy product and process specifications. We demonstrated how the integration of a mathematical algorithm that defines and quantifies the boundaries of this space can enable quantification of the space size, providing information about process behavior and its responsiveness to input variations. The capabilities of the presented framework were illustrated through an industrially relevant case study of a downstream purification unit used in biopharmaceutical manufacturing. We considered design (switching time) and operating decisions (flow rate), as well as disturbances (feed composition) simultaneously. Based on process target KPIs, performance, and feasibility constraints, we identified and quantified a DSp that includes suitable candidate condition sets predicted to meet the KPIs of interest. We further demonstrated how this space can be used for the evaluation and quantitative comparison of different operating conditions and their flexibility in handling variability arising from the operating conditions. This can assist manufacturers by offering a measurable estimate of the variability that a process at hand can handle without any further mediation while remaining within spec. Expanding this further, one could investigate the integration of the presented framework with flexibility analysis and optimization algorithms [54-58] as means to explore how the designed unit operation can impact the end-to-end process.

# Appendix A

As discussed in Section 2 of the manuscript, we employ computational geometry for the identification of the Design Space boundaries. Alpha shapes (α-shapes) are typically used to generalize bounding polygons containing sets of points. A sub-category of an α-shape is the convex hull, applicable in cases of nonlinear problems. Here we discuss the two approaches and provide details on the critical role of the choice of an appropriate alpha radius ($a_r$). We further present 3 key algorithms as developed for this framework.

*Convex Hull*

In case that the obtained point cloud containing the points satisfying both the feasibility and performance constraints, $P_{sat}$, is convex, the design space can be described using convex hulls [47-50]. Given a point cloud $P$, the convex hull of $P$ is defined as the unique minimal convex set containing $P$. For 2-dimensional point clouds, the boundary of the convex hull would be a closed curve containing $P$, with the minimum perimeter possible. For a 3-dimensional point cloud, the boundary of the convex hull would be the smallest convex bounding volume that encapsulates $P$. For this, we employ the Quickhull algorithm proposed by Barber, et al. [50], typically used for calculating the convex hull and Delaunay triangulations. In short, Barber, et al. [50] implemented an improved version of the Beneath-Beyond algorithm which is an incremental algorithm for processing the points. The algorithm has been integrated in MATLAB, GNU Octave, R, and Python.

*A.1 Algorithm A1 - Alpha Shapes*

In cases where the problem of interest contains non-linear relationships between the design decisions and the outputs of interest, the resulting point clouds may be non-convex. In such cases,

23  convex hulls may lead to over-optimistic spaces that include violated samples. Alpha shapes can
24  form convex and non-convex hulls based on a single hyperparameter; the alpha radius ($\alpha_r$). In this
25  work, we employ the algorithm proposed by Edelsbrunner and Mücke [51] based on calculating
26  Delaunay Triangulations (DT) and examining the radius of the circum-spheres of the formed
27  triangulations. A circum-sphere of a tetrahedron is a sphere for which all vertices of the tetrahedron
28  lie on the surface of the sphere. A Delaunay triangulation for a given set of points P connects the
29  points to form several tetrahedrons. The condition is that for each tetrahedron formed, the circum-
30  sphere of each tetrahedron must be empty. In this work, Algorithm A1 is used to compute the α-
31  shape.

---

Algorithm A1 Alpha shapes algorithm given an $\alpha_r$ value [51].

---

Input: Point cloud $P = \{\theta_i : i = 1,2,3,\ldots,n\}$ of size $n$ and alpha radius $\alpha_r$.

Output: Alpha shape.

1) Calculate the Delaunay triangulation of the point cloud P: $DT(P)$.
2) For all tetrahedrons in the triangulation ($DT(P)$), calculate the circum-radius, $c_r$, of the circum-sphere.
3) Eliminate tetrahedrons whose circum-radius is larger than the alpha radius $\alpha_r$.
4) Obtain the outer edges of the group of tetrahedrons. The edges are defined as triangles that are unique to the current tetrahedron (not inside of other tetrahedrons). The outer edges form the alpha shape.

---

## A.2 Algorithm A2 - The Alpha radius problem

Often, in nonlinear problems, the identification of a single α-shape is challenging, primarily due to unexplored areas that result in disjointed spaces. One way to tackle this is to identify an appropriate alpha radius ($\alpha_r$). The alpha radius influences the accuracy with which an α-shape is formed and, in this case, whether this shape is a valid representation of the Design Space [51]. Briefly, a low $\alpha_r$ value is likely to lead to an α-shape formed by >1 regions that contain no violations, while increasing values of $\alpha_r$ would increase the probability of obtaining a single shape containing points that violate one or more of the performance constraints. Figure A1 shows a simplified 2D example of the impact of the alpha radius value on an example problem.

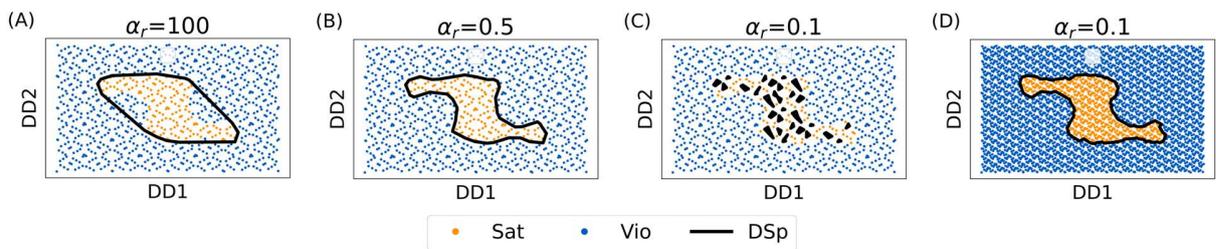

Figure A1. Illustration of the impact of alpha radius on the alpha shape of a non-convex/non-linear problem. The satisfied and violated point clouds have been classified, and alpha shapes are identified at three different alpha radius values: (A) $\alpha_r=100$, (B) $\alpha_r=0.5$, (C) $\alpha_r=0.1$, (D) $\alpha_r=0.1$ with 4 times more points than the others.

A large alpha radius value produces a convex hull as shown in Figure A1 (A) that includes violated points. As we decrease the alpha radius, the non-convexity of the problem is captured (Figure A1 (B)). However, very small values of the alpha radius can result in disjointed spaces (Figure A1 (C)). From an operational standpoint, this is undesirable, as the area of constraint satisfaction may be misrepresented.

To assist in the identification of suitable alpha radius values, we have developed Algorithm A2 which computes the alpha radius based on the parameters detailed in the *Input* section. Bisection search is used to find the largest $\alpha_r$ value that can be used while satisfying a constraint in the number of violations allowed in the alpha shape. The search is based on normalized bounds ($\alpha_{m_L}$ and $\alpha_{m_U}$) with respect to the base alpha radius value ($\alpha_b$) to increase the robustness of the search. The $v_{max\%}$ parameter is the percentage of violated points allowed to be present within the alpha shape ($v_{max\%} = \frac{P_{vio}}{P_{sat}+P_{vio}} \times 100$). The barycentric transformation strategy is implemented to identify whether a point lies inside the alpha shape or not [59].

The termination conditions of Algorithm A2 are: 1) maximum iterations have been reached, 2) the $v_{max\%}$ parameter has been satisfied and the difference between upper and lower bounds is less than the bisection tolerance. This ensures that the violation constraint ($v_{max\%}$) is satisfied for every termination except in the case of maximum iterations termination. There are only two cases that would result in a maximum iterations termination. First, is when the initial upper and lower bisection bounds are excessively large resulting in very slow steps being taken and hence a larger number of iterations needed. In that case, we can either refine the bounds or increase the maximum number of iterations. Another potential case is when the upper and lower bisection bounds are not appropriately chosen as defined in the inputs to Algorithm 2 ($v_{max\%}$ must be satisfied at lower bound, and $v_{max\%}$ must be violated at the upper bounds).

The implemented algorithm can be applied to both 2D and 3D problems. In the 2D case, the triangulations would result in triangles instead of tetrahedrons and the alpha shape boundaries would be lines instead of triangular surfaces. We have made the algorithm for 2- and 3D problems available via a python package ('*dside*') (https://github.com/stvsach/dside).

---

**Algorithm A2** Alpha radius determination algorithm using bisection method to find the largest alpha radius value which satisfies violation inside of the shape ($v_{max\%}$) parameter.

**Input**: Satisfied and violated point clouds $P_{sat}$ of size $n_{sat}$, $P_{vio}$ of size $n_{vio}$, base alpha radius value $\alpha_b = (\theta_{1_U} \times \theta_{2_U} \times \theta_{3_U})/3$ with respect to the upper bounds of the design decisions, maximum iterations $iter_{max,\alpha}$, bisection change tolerance $\Delta_{tol,\alpha}$, maximum number of violations inside the shape in percentage $v_{max\%}$, upper and lower bounds for the alpha multiplier $(\alpha_{m_U}, \alpha_{m_L})$ for which at the lower bound $v_{max\%}$ is satisfied and at the upper bound $v_{max\%}$ violated.

**Output**: Alpha shape with the determined $\alpha_r$ and number of violations inside the alpha shape.

**for** $i = 1, \dots, iter_{max,\alpha}$:

1) Current alpha multiplier (midpoint) calculation: $\alpha_m = \alpha_{m_L} + (\alpha_{m_U} - \alpha_{m_L})/2$.

2) Get alpha shape based on current alpha radius using **Algorithm$_{A1}$**:
   $\alpha_r = \alpha_b \times \alpha_m$
   $alpha\_shape = \textbf{Algorithm}_{A1}(P_{sat}, \alpha_r)$

3) Point in shape check to examine how many violated points are inside the alpha shape:
   $v_{num} = inside(P_{vio}, alpha\_shape)$

4) Calculate maximum allowable violations inside the alpha shape:
   $v_{maxn} = v_{max\%}(n_{sat} + v_{num})$

5) **if** $v_{num} \leq v_{maxn}$ **then**:
   a. Set current alpha multiplier as the lower bound: $\alpha_{m_L} = \alpha_m$
   b. **if** $(\alpha_{m_U} - \alpha_{m_L}) \leq \Delta_{tol,\alpha}$ **then**: break the **for** loop, terminate iteration.
   c. **else: pass**
   
   **else**: set current alpha multiplier as the upper bound: $\alpha_{m_U} = \alpha_m$

**end for**

---

It is worth noting that although the output of Algorithm A2 always satisfies the violation in shape ($v_{max\%}$) parameter, disjointed regions can still be formed. Hence, determining the number of

79     regions that the alpha shape is composed of allows for the identification of unwanted disjointed

80     regions. The latter refers primarily to regions that are not connected via any vertices/lines leading

81     to void areas (e.g., Figure A1 (C)). After an α-shape has been identified, a breadth-first queued

82     search algorithm is implemented [60]. The algorithm starts on one of the triangles forming the

83     alpha shape boundary and finds the neighboring triangles, sharing the same lines (pairings of

84     vertices). This is done iteratively until all triangles in the boundary have been visited and classified.

85     The number of regions formed translates into whether the formed shape is continuous or not. In

86     the case of the example problem in Figure A1, the satisfied points region is continuous not

87     separated by violated points. Algorithm A2 can be tailored to cater to the implementation of all

88     three methodologies discussed in section 2.2.

89     *A.3 Algorithm A3 - Acceptable Operating Region (AOR)*

90     As discussed in section 2.3, the framework allows the identification and quantification of an

91     AOR around a Nominal Operating Point (NOP) of interest. To do this, a bisection search is applied,

92     starting from the NOP to form a uniform polygon region (square for 2D and cube for 3D). This is

93     expanded outwards and uniformly towards all directions until one of the vertices hits the design

94     space boundary (Algorithm A3). The termination conditions of Algorithm A3 are: 1) maximum

95     iterations have been reached, 2) all vertices in $P_{verts}$ are inside the alpha shape and the difference

96     between upper and lower bounds is less than the bisection tolerance. This ensures that the formed

97     AOR will always be inside of the α-shape. Max iteration exit can mean two things: 1) the vertices

98     have not expanded enough, and all vertices are still inside the alpha shape, or 2) the bisection

99     tolerance has not been satisfied yet. For both cases, we can increase the number of iterations to

100    converge. The algorithm provided below is focused on 3D shapes. For 2D the same algorithm can

101    be implemented, the difference is that $P_{verts}$ will contain only 4 vertices which will expand

102    outwards from the NOP in four directions forming a 2D square.

**Algorithm A3** Identification of an acceptable operating region (AOR) given a 3D $\alpha$-shape and a nominal operating point (NOP).

**Input**: An alpha shape, $alpha\_shape$, nominal operating point to be investigated $p_{NOP}$, bisection tolerance, $\Delta_{tol,AOR}$, and maximum bisection iterations, $iter_{max,AOR}$.

**Output**: If $p_{NOP}$ is outside of the $alpha\_shape$, then return nothing. If it is inside, the output is the identified AOR boundary.

**if** $inside(p_{NOP}, alpha_{shape})$ **is** False **then:**

    1) Terminate the algorithm and print "$p_{NOP}$ is outside of the alpha shape".

**else**:

    1) Initialize upper and lower bound for scaled bisection, $B_U = 1$ and $B_L = 0$.

    2) Initialize a set of 8 points ($P_{verts}$) at the NOP value. This set of 8 points is the vertices of the cube that forms the AOR cantered around the NOP.

    3) **for** $i = 1, \ldots, iter_{max,DSp}$:

        i. Calculate the current fractional change for stepping within the algorithm and the bisection gap,

$$fc = B_L + \frac{B_U - B_L}{2}, B_{gap} = B_U - B_L$$

        ii. Step and update the points in $P_{verts}$ away from the NOP in all 8 directions in 3D based on the fractional change $fc$.

        iii. Check if any of the points are outside, $inside(P_{verts}, alpha\_shape)$.

        iv. **if** all $P_{verts}$ vertices are inside **then:**

            a. Set lower bisection bound to be the current fraction change, $B_L = fc$

            b. **if** $B_{gap} \leq \Delta_{tol,AOR}$ **then**:

                1. terminate and exit the algorithm with the current $P_{verts}$ as the AOR boundary.

            c. **else: pass**

        v. **else: s**et upper bisection bound to be the current fraction change, $B_U = fc$

    4) **end for**

## Appendix B

In this appendix, we present the dynamic two-column protein A affinity chromatography process for monoclonal antibody (mAb) capture [52]. More details on the mode development and parameters can be found in Steinebach, et al. [52]. Following this section, Appendix B presents the nomenclature for the model.

**Liquid phase transport through the column**

The liquid phase concentration of mAb is modeled with respect to the time and space across the length of the column. Equation (A1) shows the mass balance of the bulk liquid phase mAb concentration. The left-hand side of the equation represents the change with respect to time, while the right-hand side is the change with respect to axial diffusion, convective transport, and the adsorption phenomena.

$$\frac{\partial c(x,t)}{\partial t} = D_{ax} \frac{\partial^2 c(x,t)}{\partial x^2} - u \frac{\partial c(x,t)}{\partial x} + \frac{1-\varepsilon_b}{\varepsilon_b} \frac{3}{R_p} k_{tot} \left( c_p(x,t) - c(x,t) \right) \quad \text{(A1)}$$

**Transport through resin particle**

The solid phase concentration of mAb represents the amount of mAb adsorbed onto the protein A resin. Assuming lumped linear driving force of the transport phenomena, the mass balance is written as shown in Equation (A2). On the right-hand side of the equation, $k_{tot}$ represents the lumped mass transfer coefficient across the film surrounding the resin particle and inside the pores. The last term of the right-hand side refers to the adsorption of the mAb onto the binding sites of the protein A resin.

$$\frac{\partial c_p(x,t)}{\partial t} = \frac{3}{R_p} \frac{1}{\varepsilon_p} k_{tot} \left( c(x,t) - c_p(x,t) \right) - \left( \frac{\partial q_1}{\partial t} + \frac{\partial q_2}{\partial t} \right) \frac{1-\varepsilon_p}{\varepsilon_p} \quad \text{(A2)}$$

## Lumped mass transfer kinetics

The lumped mass transfer coefficient, $k_{tot}$, is described by two mass transfer coefficients. The film mass-transfer coefficient, $k_f$ and the transport coefficient inside the pore, $k_s$, as shown in Equation (A3).

$$\frac{1}{k_{tot}} = \frac{1}{k_f} + \frac{1}{k_s} \tag{A3}$$

Equations (A4) to (A6) details the equations used to calculate $k_f$ and $k_s$. It is noteworthy that the since the fractional coverage, $\alpha$, depends on mAb concentration in the pores, $q_1$, the value of $k_s$ is dependent on both time and spatial axes. Therefore, the value of $k_{tot}$ is also dependent on the time and spatial axes.

$$k_f = \left(\frac{D_m}{2R_p} \frac{1.09}{\varepsilon_b} \left(\frac{2uR_p}{D_m}\right)^{1/3}\right) \tag{A4}$$

$$k_s = \frac{\varepsilon_p D_p}{R_p} \frac{(1-\alpha)^{1/3}}{1-(1-\alpha)^{1/3}} \tag{A5}$$

$$\alpha = \frac{q_1}{q_{max}} \frac{1/K_a + c_{feed}}{c_{feed}} \tag{A6}$$

## Adsorption onto Protein A binding sites

A native Protein A ligand that has 5 theoretical binding sites typically results in 2-3 utilized ones. The MabSelect SuRe employs a tetramer of an alkaline stable engineered binding domain. Therefore, a dual binding site mechanism is implemented. The second site is assumed to have a slower adsorption rate and is modeled as hierarchical equilibrium adsorption. As shown in Equation (A7), the binding onto site 2 can only start after site 1 has been saturated.

$$\frac{\partial q_1}{\partial t} = k_{a1}\left[c_p(q_{max} - q_1) - \frac{q_1}{K_a}\right]$$

$$\frac{\partial q_2}{\partial t} = k_{a2}\left[c_p(q_1 - q_2) - \frac{q_2}{K_a}\right]$$

(A7)

138  A set of two columns are simulated using these equations dynamically. In addition to these
139  equations, the model includes inlet-outlet configurations between the two columns which are
140  dictated by the step of the cycle it is on. The implemented model in this work follows the same
141  equation and parameters as reported by Steinebach, et al. [52]. As shown in Figure A2, the model
142  has a perfect agreement with the reported isotherms and mass transfer coefficient with respect to
143  the saturation.

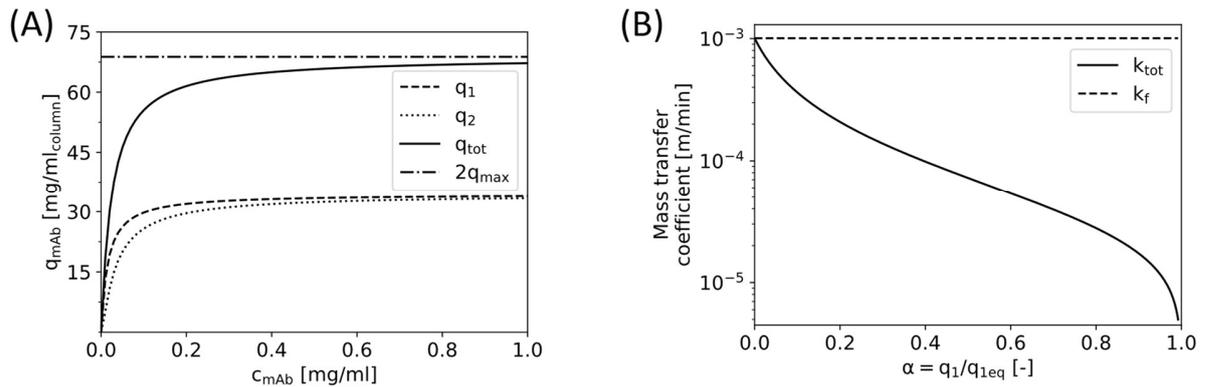

145  Figure B1. (A) adsorption isotherms of the implemented model. (B) the mass transfer coefficient with
146  respect to the saturation of the binding sites. Both agree perfectly with the ones reported by Steinebach, et
147  al. [52].

148

149 **Appendix C**

150 Table B1 details the violated points lying inside of the combinatorial method alpha shapes shown
151 in Figure 8 (H) – (J).

152 Table C1. Details on the violated points lying inside of the combinatorial method alpha shapes. The
153 Unique ID identifies the unique point which may be shared across the identified alpha shapes. Alpha
154 shape ID is with respect to Figure 8 (H) – (J). The tolerance used is 0.25% allowed violation points inside
155 the shape.

| Unique ID | Alpha Shape ID | No Extra Points | $c_{feed}$ (mg ml$^{-1}$) | $Q_{feed}$ (ml min$^{-1}$) | $T_{switch}$ (min) | Yield (%) | Productivity (mg ml$^{-1}$ h$^{-1}$) |
|---|---|---|---|---|---|---|---|
| 1 | H | 1024 | 0.32 | 0.93 | 99.7 | 98.99 | 5.41 |
| 2 | H | 1024 | 0.45 | 0.81 | 87.1 | 98.86 | 6.64 |
| 3 | H | 1024 | 0.42 | 1.00 | 61.6 | 98.93 | 7.67 |
| 4 | I | 2048 | 0.30 | 0.99 | 96.9 | 98.99 | 5.36 |
| 1 | I | 2048 | 0.32 | 0.93 | 99.7 | 98.99 | 5.41 |
| 5 | I | 2048 | 0.42 | 0.81 | 101.0 | 98.64 | 6.12 |
| 1 | J | 4096 | 0.32 | 0.93 | 99.7 | 98.99 | 5.41 |
| 2 | J | 4096 | 0.45 | 0.81 | 87.1 | 98.86 | 6.64 |
| 5 | J | 4096 | 0.42 | 0.81 | 101.0 | 98.64 | 6.12 |

156

157 The spaces defined in Figure 8 (H), (I), and (J) use 1024, 2048, and 4096 extra points
158 respectively. Some of the violation points are shared within the three different alpha shapes. All
159 three alpha shapes share point no 1. While shape I and J share point 5, shape H and J share point
160 2. The maximum error in the yield is in point 5 (0.36%), present in shapes J and I.

# Appendix D

## Nomenclature

*Abbreviations*

| | | |
|---|---|---|
| 164 | ANN | artificial neural network |
| 165 | AOR | acceptable operating region |
| 166 | CMA | critical material attributes |
| 167 | Comb. | combinatorial design space identification method |
| 168 | CPP | critical process parameter |
| 169 | CQA | critical quality attribute |
| 170 | DD | design decision |
| 171 | DoE | design of experiment |
| 172 | DSp | design space |
| 173 | HIC | hydrophobic interaction chromatography |
| 174 | KPI | key performance indicator |
| 175 | KSp | knowledge space |
| 176 | MPAR | multivariate proven acceptable range |
| 177 | NOP | nominal operating point |
| 178 | ODE | ordinary differential equation |
| 179 | PAR | proven acceptable range |
| 180 | PAT | process analytical tool |
| 181 | PCC | periodic counter-current (chromatography process) |
| 182 | PDAE | partial differential and algebraic equation |
| 183 | QbD | quality by design |
| 184 | QbDD | quality by digital design |
| 185 | QbT | quality by testing |
| 186 | RS | resolution support design space identification method |
| 187 | Tol. | tolerance-based design space identification method |
| 188 | TPP | target product profile |
| 189 | ***Physical and Mathematical Quantities*** | |
| 190 | $alpha\_shape$ | alpha shape of a point cloud |
| 191 | $B_{gap}$ | current bisection gap in AOR identification in Algorithm A3 |
| 192 | $B_L, B_U$ | lower and upper bounds in Algorithm A3 for AOR identification |
| 193 | $c$ | concentration of mAb in the bulk liquid phase, mg/ml |
| 194 | $c_{feed}$ | monoclonal antibody concentration in the feed to the process |
| 195 | $c_p$ | concentration of mAb in the pores, mg/ml |
| 196 | $D_{ax}$ | axial diffusion coefficient, cm²/s |
| 197 | $f$ | process model |
| 198 | $fc$ | fraction change for stepping outwards from the NOP in the AOR identification |
| 199 | **$g$** | vector of performance constraints |
| 200 | $iter_{max,\alpha}$ | maximum iterations of the alpha radius problem in Algorithm A2 |
| 201 | $iter_{max,AOR}$ | maximum iterations of the AOR identification in Algorithm A3 |
| 202 | $K_a$ | equilibrium constant of adsorption process, ml/mg |
| 203 | $k_{a1}, k_{a2}$ | adsorption rate constant for site 1 and 2, ml/mg/min |
| 204 | $k_f$ | film mass transfer coefficient, cm/s |
| 205 | $k_s$ | inside pore mass transfer coefficient, cm/s |
| 206 | $k_{tot}$ | lumped mass transfer coefficient, cm/s |
| 207 | $n_p$ | number of inputs: $2^{sp}$ |
| 208 | $n_{sat}, n_{vio}$ | number of satisfied points and violated points |
| 209 | $n_\theta$ | size of the design decisions vector |
| 210 | $P_{KSp}$ | point cloud of the knowledge space |
| 211 | $p_{NOP}$ | investigated nominal operating point in Algorithm A3 |
| 212 | $P_{sat}$ | point cloud of the satisfied points |
| 213 | $P_{sat,extra}$ | extra points obtained from the neural network interpolation |
| 214 | $P_{verts}$ | vertices of the cube that makes up the AOR in Algorithm A3 |
| 215 | $P_{vio}$ | point cloud of the violated points |
| 216 | $Q_{feed}$ | volumetric flowrate of the feed to the process |
| 217 | $q_1, q_2$ | concentration of mAb on binding site 1 and 2, mg/ml |
| 218 | $R_p$ | resin particle radius, cm |
| 219 | $sp$ | Sobol sequence number of samples is based on 2 to the power of $sp$ |
| 220 | $T_{switch}$ | column switching time in the multicolumn protein A chromatography process. |
| 221 | $u$ | column liquid velocity, cm/s |
| 222 | $v_{max\%}$ | maximum number of violated points allowed inside of the shape in percentage |
| 223 | $v_{maxn}$ | number of violation points allowed inside the shape |
| 224 | $v_{num}$ | number of violations inside the shape |
| 225 | $y$ | vector of monitored key performance indicators |
| 226 | $\alpha$ | fractional coverage, dimensionless |
| 227 | $\alpha_b$ | base alpha radius |
| 228 | $\alpha_m$ | alpha multiplier for alpha radius determination via bisection |
| 229 | $\alpha_{m_L}, \alpha_{m_U}$ | lower and upper bisection bounds for alpha multiplier |
| 230 | $\alpha_r$ | alpha radius |
| 231 | $\Delta_{tol,\alpha}$ | bisection tolerance of the alpha radius problem in Algorithm A2 |
| 232 | $\Delta_{tol,AOR}$ | bisection tolerance of the AOR identification in Algorithm A3 |
| 233 | $\varepsilon_b$ | bed porosity, dimensionless |
| 234 | $\varepsilon_p$ | resin particle porosity, dimensionless |
| 235 | $\boldsymbol{\theta}$ | vector of design decisions |
| 236 | $\theta_{in}$ | set of design decisions generated based on the Sobol sequence |
| 237 | $\boldsymbol{\theta}_L, \boldsymbol{\theta}_U$ | vector of design decisions feasibility constraint (lower and upper bounds) |